\documentclass[aps,twocolumn,preprintnumbers,amsmath,amssymb]{revtex4}
\usepackage{dcolumn}
\usepackage{bm}
\usepackage{graphicx,subfigure,xcolor}
\usepackage{braket}
\usepackage{bibunits}
\usepackage{xcolor}
\DeclareMathOperator{\Tr}{Tr}

\newcommand{\bd}{b^{\dag}}

\begin{document}
\begin{bibunit}

\title{Quantum and classical phases in optomechanics}

\author{Federico Armata$^1$}

\author{Ludovico Latmiral$^1$}

\author{Igor Pikovski$^{2,3}$}

\author{Michael R. Vanner$^4$}

\author{\v{C}aslav Brukner$^{5,6}$}

\author{M. S.  Kim$^1$}

\affiliation{$^1$QOLS, Blackett Laboratory, Imperial College London, London SW7 2AZ, United Kingdom \\ 
$^2$ITAMP, Harvard-Smithsonian Center for Astrophysics, Cambridge, MA 02138, USA \\
$^3$Department of Physics, Harvard University, Cambridge, MA 02138, USA \\
$^4$Clarendon Laboratory, Department of Physics, University of Oxford, OX1 3PU, United Kingdom \\
$^5$Faculty of Physics, University of Vienna, Boltzmanngasse 5, 1090 Vienna, Austria\\
$^6$Institute for Quantum Optics and Quantum Information, Boltzmanngasse 3, 1090 Vienna, Austria
}

\begin{abstract}
The control of quantum systems requires the ability to change and read-out the phase of a system. The non-commutativity of canonical conjugate operators can induce phases on quantum systems, which can be employed for implementing phase gates and for precision measurements. Here we study the phase acquired by a radiation field after its radiation pressure interaction with a mechanical oscillator, and compare the classical and quantum contributions. The classical description can reproduce the nonlinearity induced by the mechanical oscillator and the loss of correlations between mechanics and optical field at certain interaction times. Such features alone are therefore insufficient for probing the quantum nature of the interaction. Our results thus isolate genuine quantum contributions of the optomechanical interaction that could be probed in current experiments.
\end{abstract}

\date{\today}

\maketitle

\section{\label{sec:level1}Introduction}

In quantum optics, single and multimode nonlinearities are of use for quantum information processing tasks and for tests of foundational physics. Such nonlinearities can be generated from more readily available (linear) operations by enclosing loops in phase or parameter space. This technique is now utilized broadly throughout both theoretical and experimental quantum science with one prominent example being trapped-ion systems \cite{milburn2000, sorensen2000, Leibfried2003}. A set of simple operations is applied in a sequence to enclose such a loop and deterministically generate an effective non-linearity on the other system. These types of nonlinearities and phases, which have a geometric interpretation, are also very valuable in optical~\cite{Kwiat2005, Langford2011} and superconducting circuit~\cite{Leek2007} experiments.

Quantum optomechanics, which exploits the radiation-pressure interaction between an optical field and a mechanical element~\cite{aspelmeyer2014}, is an emerging area of quantum science that is now gaining increasing interest in such nonlinearities. A key goal of the field is to explore non-classical properties of mechanical motion, which can be generated by enclosing loops in the phase space of either the optical~\cite{khosla2013} or mechanical degrees~\cite{pikovski2012, latmiral2016} of freedom. Indeed, the seminal works of Refs. \cite{mancini1997, bose1997}, which study a continuous interaction between an optical cavity field and a mechanical element, have an implicit closed loop in the dynamics where the mechanical oscillator undergoes a closed pattern in phase space and the optical field picks up a nonlinear phase. Bose {\it et al.} \cite{bose1999} noticed that at a certain interaction time, the optical field state decouples from the oscillator state and proposed to leverage this peculiarity for decoherence sensing. This idea was further developed by Armour {\it et al.} in Ref.\cite{armour2002}, where a micromechanical resonator is capacitively coupled to a Cooper-pair-box and then, by Marshall {\it et al.} in Ref.\cite{penrose2003}, where correlations between a single-photon path-entangled optical state and a mechanical object are used to study gravitational decoherence mechanisms \cite{penrose1998}. In the latter scheme, the interference visibility between the two components of the optical field is used as a witness of mechanical coherence, which can be degraded by both standard decoherence and potential gravitational collapse mechanisms. A key to their proposal is observing a recovery of the interference visibility, which arises when the light-mechanics system becomes disentangled after the mechanics completes a closed loop in phase space. Other optomechanical proposals consider a nonlinear phase imparted on a qubit after the mechanical oscillator undergoes a closed loop \cite{vacanti2012, scala2013}.

The optomechanical interaction has been studied extensively in quantum mechanics. In this paper, we analyze the non-classicality of optomechanical phases by studying the dynamics in a fully classical picture and comparing it with the quantum prediction. We provide a general mathematical framework and focus our discussion on two proposals: Ref.~\cite{pikovski2012} and Ref.~\cite{penrose2003}. We start by considering the \emph{pulsed} interaction regime \cite{braginsky1995, vanner2011} of Ref.~\cite{pikovski2012} and then discuss the \emph{continuous} interaction regime of Ref.~\cite{penrose2003} through which we explore the evolution of interferometric visibility. We find that many of the features which have been tacitly considered quantum signatures in such setups can be reproduced classically. Specifically, in the context of the pulsed regime discussed in Ref.\cite{pikovski2012}, we prove that a large amount of the quantum phase has a classical nature. We also find that the main peculiarities of the quantum phase have a correspondence in classical physics: the nonlinearity induced by the mechanical oscillator and its independence of the oscillator state at some interaction times. Surprisingly, this is key to prove that the loss and revival of the visibility pattern in the interferometric scheme discussed in Ref.~\cite{penrose2003}, which have been considered a quantum signature of the system dynamics, can be explained by a completely classical description of the model. On the other hand, we are able to identify nonclassical components to the dynamics that cannot be obtained classically or semi-classically.

\section{\label{sec:level2} The Model}

\begin{figure}[t!]
\centering
\includegraphics[scale=0.23]{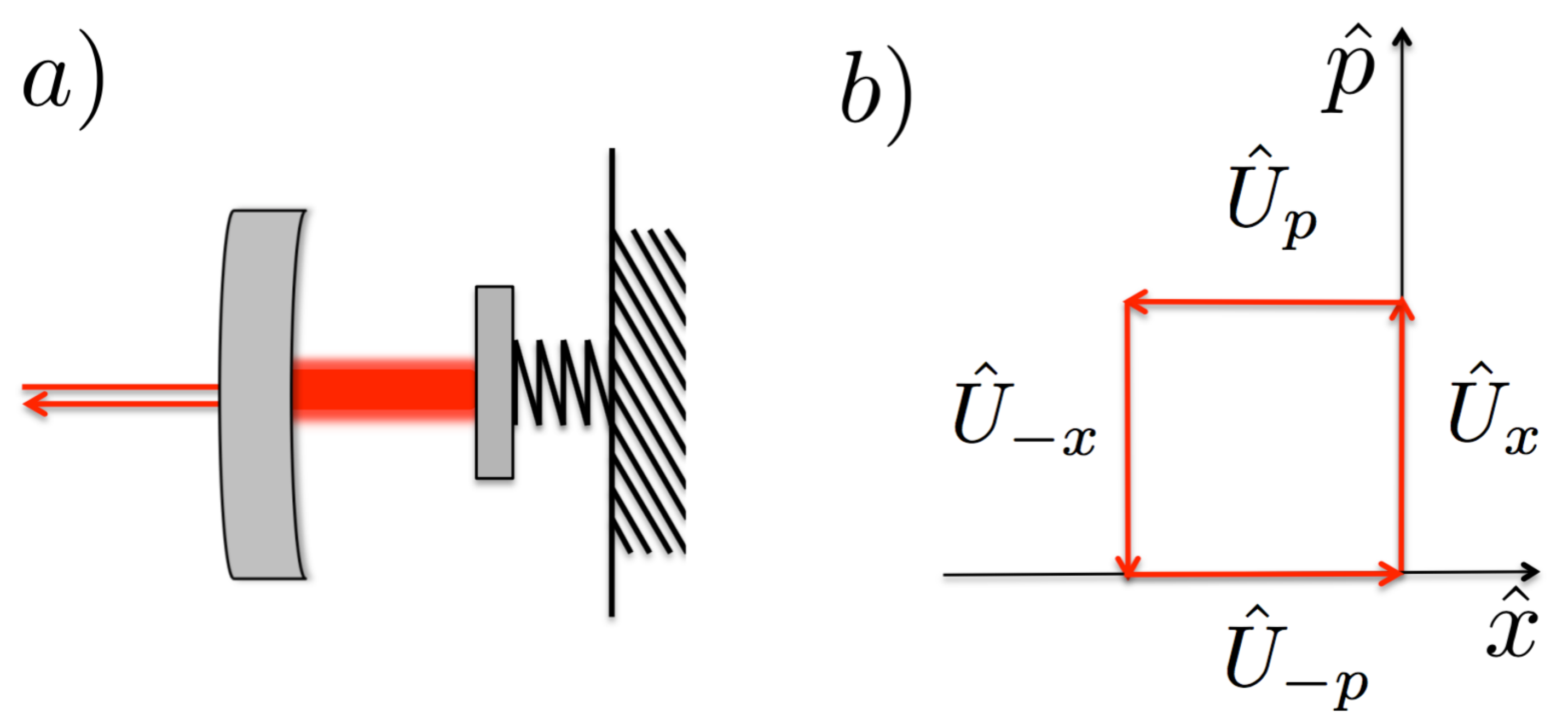}
\caption{a) Optomechanical cavity with a harmonically oscillating mirror at one end. b) Scheme of a four displacement operation in the phase space of the mechanical oscillator.}\label{kicks}
\end{figure}

We consider a mechanically oscillating mirror of frequency $\omega$ and mass $m$ coupled to an optical field of frequency $\omega_f$ inside a cavity of mean length $L$ [see Fig.\ref{kicks}(a)]. The effective Hamiltonian that describes this system in a frame rotating with the field can be written as \cite{pace1993, law1994}
\begin{equation}
\hat{H}=\hat{H}_0 - \hbar g_0\hat{n}\hat{x},
\label{HOpt}
\end{equation}
where $\hat{H}_0=\frac{\hbar\omega}{2}(\hat{x}^2+\hat{p}^2)$ represents the mechanical free energy, $\hat{n}$ is the number operator of the optical field, $\hat{x}=(1/\sqrt{2})(\hat{b}^{\dag}+\hat{b})$ and $\hat{p}=(i/\sqrt{2})(\hat{b}^{\dag}-\hat{b})$ are the mirror quadrature operators and $g_0 = \omega_{f} x_0 / L$ is the optomechanical coupling rate for $x_0 = \sqrt{\hbar/m\omega}$. In the case of the short pulsed regime, the interaction time is much smaller than a period of mechanical motion $\tau=2\pi/\omega$ and the system operates in the bad cavity limit $\kappa\gg\omega$ where $\kappa$ is the cavity amplitude decay rate. We also require the characteristic mechanical decoherence time to be lower than the mechanical period. In such a regime, we can neglect the mechanical free evolution during the light-mirror interaction. The dynamics can thus be described using the unitary evolution operator $\hat{U}_x=e^{i\lambda \hat{n}\hat{x}}$ \cite{vanner2011}, where $\lambda=g_0/\kappa$ is the dimensionless coupling strength. As in Ref.~\cite{pikovski2012} we now consider a sequence of four interactions with the same pulse, each interaction being separated by a quarter of a period of mechanical motion. We write this procedure as $\hat{U}_x$, followed by $\hat{U}_p$, $\hat{U}_{-x}$, and $\hat{U}_{-p}$. This sequence of four pulses generates a square loop in mechanical phase space [see Fig.\ref{kicks}(b)] with a photon-number-dependent side-length. The net interaction of the sequence can be described by the unitary $\hat{\xi} =e^{i\lambda^2\hat{n}^2}$. This effective interaction is a highly nonlinear self-Kerr interaction. This type of nonlinearity is central in Ref.~\cite{khosla2013} and is closely analogous to the controlled gate operations in trapped ion qubits using the phononic mode of the harmonic oscillator as a mediator \cite{milburn2000, sorensen2000, Leibfried2003}.

\section{\label{sec:level3 } Quantum vs classical dynamics}

To compute the dynamics predicted quantum mechanically, we take the field initially in a coherent state $\ket{\alpha}_f$ and the mirror in an arbitrary initial state. We apply the four-pulse operator $\hat{\xi}$ and we compute the mean value of the optical field $\langle \hat{a}\rangle=\alpha\; e^{-N_{p}(1-\cos 2\lambda^2)} e^{i(\lambda^2+N_{p}\sin 2\lambda^2)}$, where $N_p=|\alpha|^2$ is the mean photon number. We observe that both modulus and phase are changed by the non-linear interaction. While the magnitude of this expectation value is reduced due to the coherent state spreading out in phase-space [for small coupling it scales as $O(e^{-N_p\lambda^4})$], the mean phase shift results to be
\begin{equation}\label{Opticalphase-t}
\varphi_{q}=(\lambda^2+N_{p}\sin 2\lambda^2).
\end{equation}
The fact that Eq.\eqref{Opticalphase-t} is independent of the state of the mirror and depends on the intensity of the optical field is going to play a crucial role for the forthcoming considerations. 

Since the phase derives from commutation rules both of field and oscillator, we would like to explore its derivation from a fully classical perspective without invoking quantum operators in order to see to what extent it can be considered an indicator of non-classicality.\\
The phase associated with a single reflection of a field on a movable mirror is proportional to the product of the field wavevector $\mathrm{k}_f$ times the mirror position \cite{rakhmanov2001, rakhmanov2002}. A radiation-pressure kick, i.e. a pulse that transfers a momentum $\mathcal{I}$ to the mechanical oscillator, can be classically depicted as the sum of $N_{rt}$ round trips of the light inside the cavity. In the pulsed regime, where the position of the movable mirror is essentially fixed during the $N_{rt}$ reflections, we imagine that the optical field enters the cavity, escapes after a time equal to the inverse of the decay rate $1/\kappa=(2L/c)N_{rt}$ and then waits in an engineered loop before being injected again. During the time between two consecutive kicks the movable mirror freely evolves accordingly to the equation of motion of a harmonic oscillator. For every radiation-pressure kick the field picks up an additional phase due to the movable mirror dynamics. We eventually have a net classical phase of 
\begin{equation}
\varphi_{c}=2\mathrm{k}_fN_{rt}\sum_{j=0}^{3}x(t_j),
\end{equation}
where $x(t_j)$ are the positions of the mirror at times $t_j=j\tau/4$.
Solving the equations of motion (see Appendix \ref{appendix1} for more details), we obtain for the classical phase
\begin{equation}\label{classical-harmonic-phase-t}
\varphi_{c}=4 \mathrm{k}_f N_{rt}\frac{\mathcal{I}}{m\omega},
\end{equation}
which linearly depends on the light intensity as described by a (classical) nonlinear Kerr effect. We also note that the classical phase does not depend on the initial conditions of the mechanical oscillator. Hence, the two features which are at the heart of the quantum operations recur also in the classical picture. In order to quantitatively compare Eq. \eqref{classical-harmonic-phase-t} with the quantum-mechanical prediction, we substitute the characteristic parameters of the optomechanical system and the transferred momentum $\mathcal{I}=2\mathrm{k}_fN_{rt}\hbar N_p$, obtaining $\varphi_{c}=2\lambda^2N_{p}$. We therefore find that quantum and classical predictions for the optical phase shift generally differ, though for small coupling strengths this difference is mainly in the form of a (small) offset $\lambda^2$.

Reducing the waiting time between subsequent pulses, it is possible to generalize the argument to $\mathcal{N}$ kicks, where loops in the shape of $\mathcal{N}$-sided polygons are enclosed in mechanical phase space \cite{vacanti2012} (see Appendix \ref{appendix1} for more details). The limit $\mathcal{N}\rightarrow\infty$ coincides with the continuous dynamics, i.e. when light remains in the cavity for the entire mechanical period. Even though it is impossible to tune the same experimental apparatus to achieve this limit, still theoretically we can correctly recover a continuous dynamics from a pulsed regime. By explicitly solving the quantum dynamics and tracing out the mechanical degrees of freedom, we find the reduced density matrix of the field $\hat{\rho}_{f}$ and the mean value of the optical field $\langle \hat{a}\rangle=\Tr[\hat{a}\hat{\rho}_{f}]$. The resulting optical phase shift for a closed loop is $\varphi_q=2\pi k^2+N_p\sin[4\pi k^2]$ where $k=g_0/(\sqrt{2}\omega)$ is the ratio between the single photon optomechanical coupling rate and the mechanical resonance frequency (see Appendix \ref{appendix2} for more details on the derivation of the quantum continuos phase). On the other hand, from a classical perspective the continuous interaction can be depicted as a constant force during the whole evolution, whose intensity is given by the field energy $E_0$. The classical Hamiltonian will then be
\begin{equation}\label{classical-hamiltonian}
H_c=\frac{1}{2}m\omega^2x^2+\frac{p^2}{2m}-\frac{E_0}{L}x,
\end{equation}
and the classical phase can be accordingly generalized to the integral over mirror positions as
\begin{equation}
\varphi_c(\tau)=2\frac{\mathrm{k}_f}{d\tilde{\tau}}\int_0^{\tau}x(\tau')d\tau',
\end{equation}
with $d\tilde{\tau}=2L/c$ the single round trip time. By working out the classical continuous dynamics, the value of the phase in the case of a closed loop results $\varphi_{c}=2\pi\omega_fE_0/(\omega^3mL^2)$ and substituting the optomechanical parameters it reads $\varphi_{c}=4\pi k^2N_p$ (see Appendix \ref{appendix2} for more details on the derivation of the classical continuous phase).

When comparing the phase predicted quantum mechanically $\varphi_q$ to the classical case $\varphi_c$ the main difference is an offset, which is equal to $\lambda^2$ for the four pulse case and depends on $k^2$ for the continuous case. We highlight that this offset is not predicted with semi-classical descriptions where either the light or mechanics are quantized and the other is treated classically (see Appendix \ref{appendix3} for further details on the semiclassical model). Experimentally observing this offset would therefore demonstrate the quantum nature of the interaction between the light and the mechanics. Such an evidence could be provided by measuring the phase as a function of the photon number per pulse $N_p$ and fitting the resultant data to obtain an estimate for the offset. Counter-intuitively, we remark that a large optomechanical coupling is not strictly necessary for the purpose, as long as the phase can be measured with a high precision. Indeed, uncertainty is mainly amenable to the quantum noise of the coherent state probe, which scales approximately as $\delta\varphi_q\sim1/\sqrt{N_p N_r}$, where $N_r$ is the number of averages. We thus require $\delta\varphi_q<\lambda^2$ to provide a good estimate for the quantum offset, which can be easily achieved with current experiments ($10^{-5}\lesssim\lambda\lesssim10^{-1}$ and $N_p\sim10^8$) \cite{brennecke2008,leijssen2015}.

Aside from this small phase shift that certifies the quantum nature of the interaction, we pinpoint that, in the context of pulsed interactions, the non-linear phase of the optical field is mainly due to the classical contribution. If the quantum nature of the system is relevant for the interpretation of an experiment, such as in Ref. \cite{pikovski2012}, it might be necessary, in order to verify the non-classical nature of the interaction, to rely on quantum state preparation of the mechanics, to study the non-classical photon statistics after the interaction or to observe the quantum offset discussed above.

\section{\label{sec:level4} Interferometer visibilities}

\begin{figure}[t!]
\centering
\subfigure[]{
        \label{interferometer}
        \includegraphics[width=0.34\textwidth]{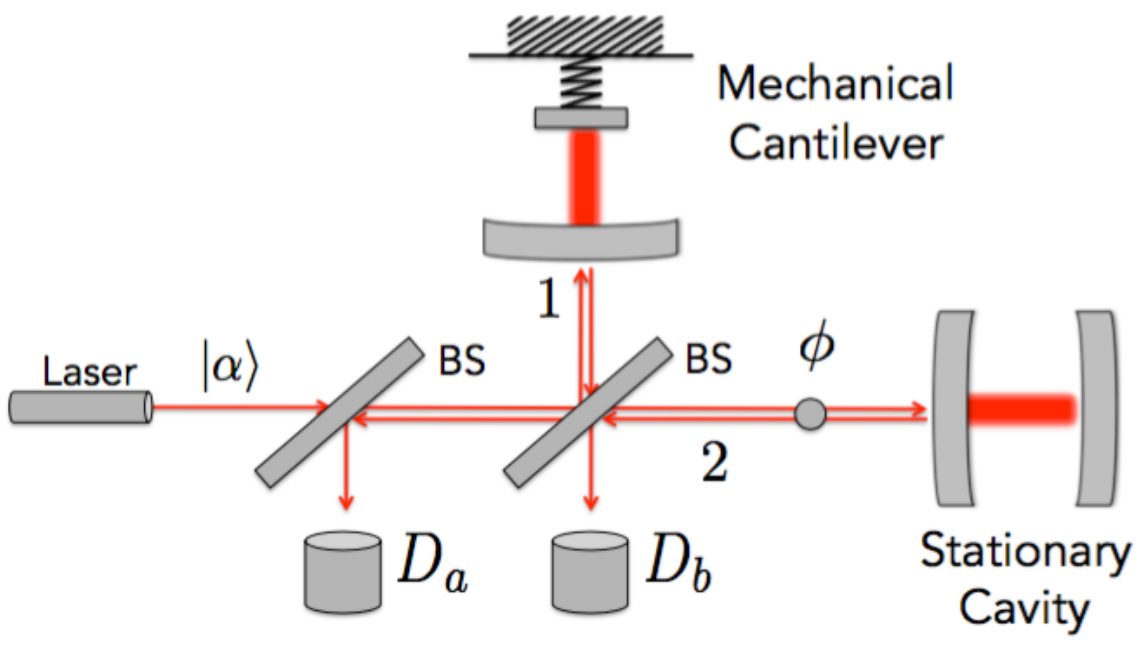} } 
\subfigure[]{
        \label{visibility_quantum}
        \includegraphics[width=0.34\textwidth]{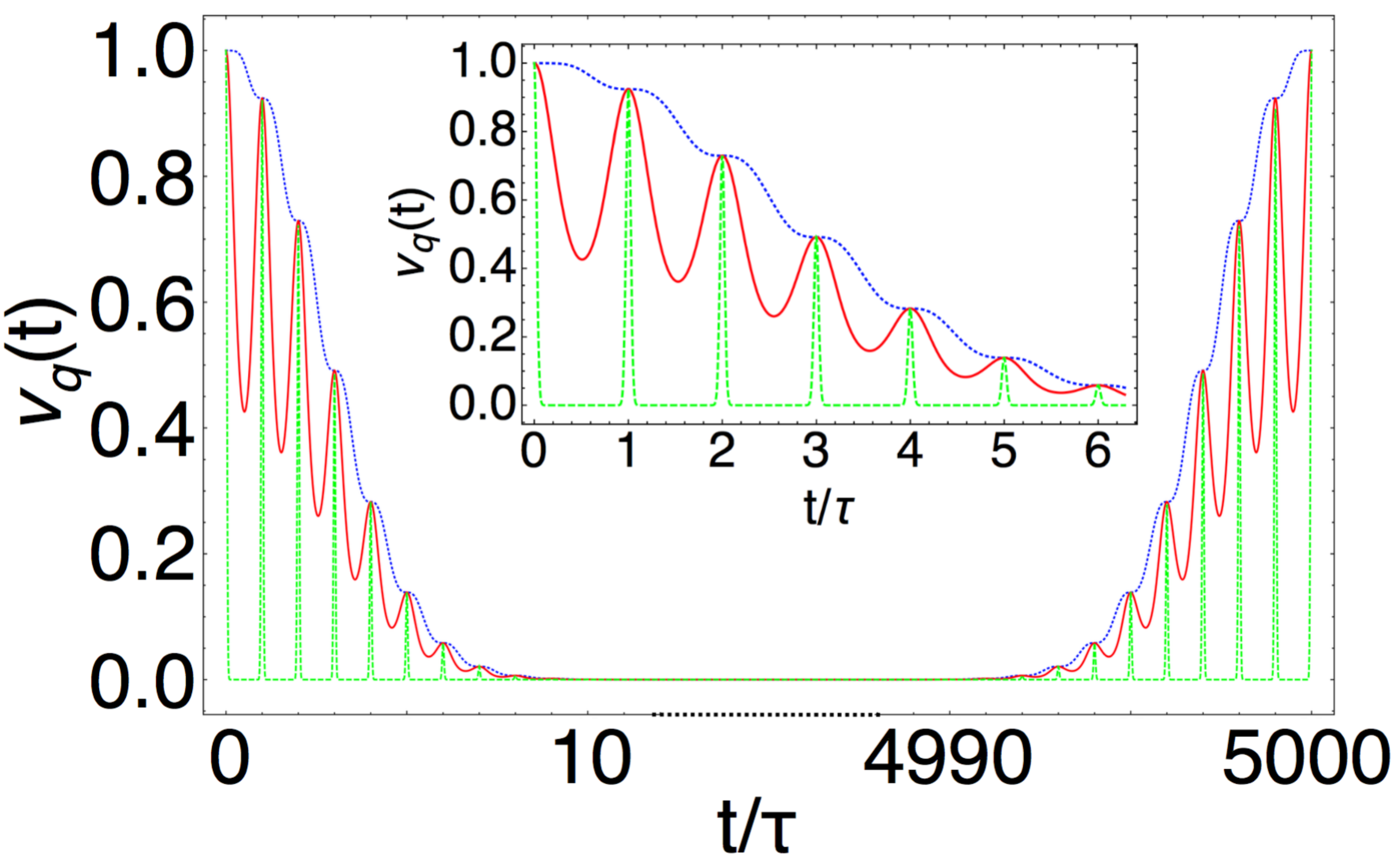} } 
\subfigure[]{
        \label{visibility_qc}
        \includegraphics[width=0.34\textwidth]{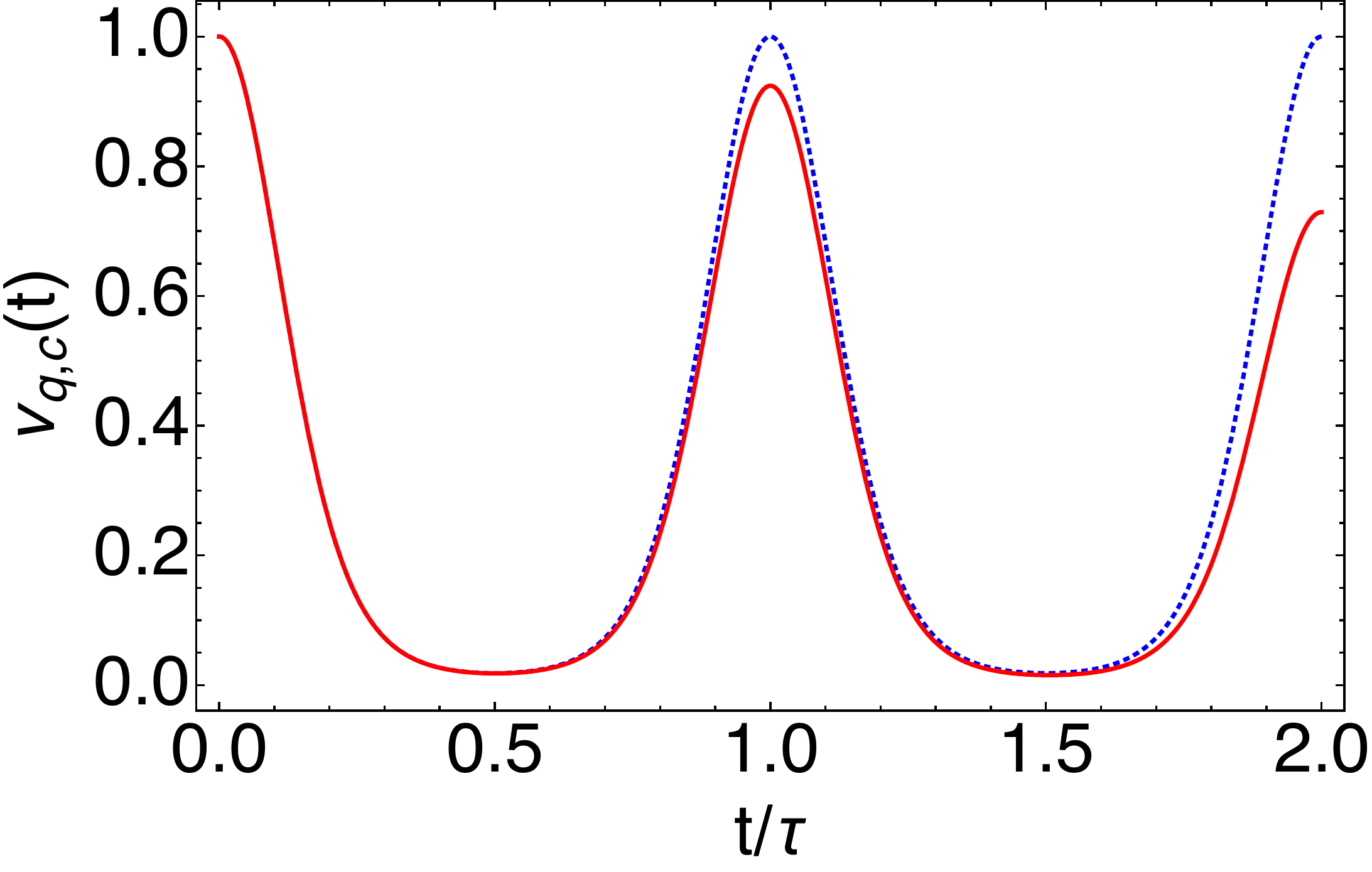}}     
\caption{a) Michelson interferometer: a coherent field $\ket\alpha_f$ is split by a beam splitter (BS) in the two arms of the interferometer. Arm 1 ends with an optomechanical cavity with a movable oscillator, while arm 2 is composed by a phase shifter and a stationary cavity.
b) Quantum visibilities $\nu_q(t)$ in Eq.\eqref{visibility} for $T=10^{-5}$K (blue dotted line), $T=10^{-2}$K (red continuous line) and $T=1\;$K (green dashed line); optomechanical coupling $k=10^{-2}$, number of photons $N_p=10^5$ and period $\tau=10^{-5}s$. For relatively high temperature the visibility is strongly suppressed within every single oscillating period. Instead, in the low temperature limit visibility is slightly lowered and the main effect is due to the Kerr non-linearity experienced by the field.
c) Quantum (red continuous line) and classical (blue dotted line) visibilities in Eqs. \eqref{visibility} and \eqref{Vis-Full-Class-t} for temperature $T=5\times 10^{-2}$K. Other parameters are as in b).}
\label{Fig2}
\end{figure}

We have observed that for closed loops in both classical and quantum pictures the phase does not depend on the initial conditions of the mechanical oscillator: we will see how this property has a non-trivial implication on the quantum-classical comparison.
Consider the Michelson interferometer depicted in Fig.~\ref{interferometer} where the end mirror of the cavity in arm $1$ interacts with an incoming coherent state via the Hamiltonian in Eq. \eqref{HOpt}. We first compute the quantum dynamics and assume the mirror initially prepared in a thermal state $\hat{\rho}_m=(\pi \bar{n})^{-1}\int d^2\gamma\; e^{-|\gamma|^2/\bar{n}}\ket{\gamma}_m\bra{\gamma}$, where $\bar{n}=1/(e^{\beta\hbar\omega}-1)$ is the average thermal occupation number and $\beta=(k_BT)^{-1}$.
By solving the Liouville equation $\dot{\hat{\rho}} = -(i/\hbar)[\hat{\rho},\hat{H}]$ for the system density matrix and tracing out the mechanical degrees of freedom, it is possible to recover the reduced density matrix of the field that allows us to calculate the light intensities on detectors $D_a$ and $D_b$.
Defining $I_{max}$ $(I_{min})$ as the maximum (minimum) intensity on the detectors, the visibility is given by the ratio $\nu =(I_{max}-I_{min})/(I_{max}+I_{min})$, which can be written conveniently as $\nu_q(t)=\nu_q^{cor}(t)\nu_q^{Kerr}(t)$ where 
\begin{eqnarray}\label{visibility}
\nu_{q}^{cor}(t)&=&e^{-k^2(1-\cos\omega t)(2\bar{n}+1)} \nonumber \\ 
\nu_q^{Kerr}(t)&=&e^{-N_{p}[1-\cos(2k^2(\omega t-\sin\omega t))]}.
\end{eqnarray}
(see Appendix \ref{appendix4} for more details on the derivation of Eq.\eqref{visibility}). As shown in Fig.~\ref{visibility_quantum}, the visibility is the composition of two periodic functions with different frequencies that settle two time scales, being responsible for two revivals. The short one of period $\tau$ is due to the term $\nu_q^{cor}(t)$, while the long one of $\tau'=\tau/(2k^2)$ is related to $\nu_q^{Kerr}(t)$. For the former, the revivals of the visibility are explained by the decoupling of field and mirror after periods of the mechanical evolution (i.e. for closed loops in phase space). This demonstrates the presence of correlations between field and mirror at intermediate times. These revivals are clearly manifested in Fig.~\ref{visibility_quantum}. On the other hand, $\nu_q^{Kerr}(t)$ is due to the Kerr non-linear interaction experienced by the field when entering into the cavity because of Hamiltonian \eqref{HOpt} and it is responsible for a reduction of the interferometric pattern. In other words, this reduction of visibility stems from the onset of squeezing of the coherent optical state due to the Kerr nonlinearity. As a result, even if mirror and field are completely uncorrelated after an interaction that lasts a mechanical period, we still cannot fully recover visibility. 

We are now going to show how the visibility recovery can be explained through a fully classical treatment, thanks to the periodic restoration of phase independence from the initial mechanical conditions. We assume the mirror initially subjected to classical thermal fluctuations around the origin described by a Maxwell-Boltzmann distribution. By using polar coordinates $(\theta,\varrho)$ we define the initial position and momentum of the oscillator as
\begin{equation}\begin{split}
&x(t=0,\;\theta,T)=\sqrt{2/(m\omega^2)}\varrho(T)\cos\theta\\
&p(t=0,\;\theta,T)=\sqrt{2m}\varrho(T)\sin\theta\\
&\varrho^2(T)=\frac{m\omega^2}{2}x^2(0,\theta,T)+\frac{p^2(0,\theta,T)}{2m},
\end{split}\end{equation}
with $\varrho^2(T)$ the initial thermal energy of the oscillator at temperature $T$. The phase acquired by the field after an interaction time $t$ will consequently depend on these initial conditions
\begin{equation}\begin{split}\label{classphase-t}
\varphi_{c}(\varrho,\theta,t)=&\sqrt{2}\chi\varrho\left[ \cos\theta\sin\omega t+\sin\theta(1-\cos\omega t)\right]\\
&+\frac{\omega}{\omega_f}\chi^2E_0\left(\omega t-\sin\omega t\right),
\end{split}\end{equation}
where $\chi=\omega_f/(\omega^2L\sqrt{m})$, (see Appendix \ref{appendix2} for more details on the derivation of the classical continuous phase). If we set with $I_0$ the intensity of the incoming field, the intensities on the detectors $D_a$ and $D_b$ depend on the phase between the two arms as
\begin{equation}
I^a_b(\varrho,\theta,t)=\frac{I_0}{2}[1 \pm \cos(\varphi_{c}(\varrho,\theta,t)-\phi)].
\end{equation}
By averaging over all initial mechanical states, we thus obtain
\begin{equation}\label{class-int-ab}\begin{split}
\langle I^a_b(t)\rangle&=\frac{\beta}{\pi}\iint \varrho\;d\varrho\ d\theta\; I^a_b(\varrho,\theta,t)\;e^{-\beta\varrho^2}\\
&=\frac{I_0}{2}\bigg[1 \pm e^{-\frac{\chi^2}{\beta}(1-\cos\omega t)}\\
&\hspace{1cm}\times\cos\left(\frac{\omega}{\omega_f}E_0\chi^2(\omega t-\sin\omega t)-\phi\right)\bigg].
\end{split}\end{equation}
It is then possible to derive the expression for the classical visibility by maximizing and minimizing Eq.\eqref{class-int-ab} 
\begin{equation}\begin{split}\label{Vis-Full-Class-t}
\nu_{c}(t)=e^{-\frac{\chi^2}{\beta}(1-\cos\omega t)},
\end{split}\end{equation}
which reveals a fully classical revival after each period of the mechanical oscillator. These revivals are due to the particular property of the phase acquired by the field that still holds in the classical scenario, i.e. its independence from the initial mirror conditions after periods of the mechanical oscillator. Indeed, the loss of visibility has to be attributed to the uncertainty on the initial conditions due to the thermal fluctuations of the mirror, which appear in the same form both in quantum and classical pictures. In contrast, in the case of zero temperature, the classical visibility will result equal to one at all times. 

Let us now compare the classical result in Eq. \eqref{Vis-Full-Class-t} with the fully quantum one $\nu_q$. By using the optomechanical parameters we get $k=\chi\sqrt{\hbar\omega/2}$ and hence the classical visibility can be written as $\nu_c(t)=e^{-\frac{2k^2}{\beta\hbar\omega}(1-\cos\omega t)}$. First of all, we confirm that the quantum thermal part of the visibility $\nu_q^{cor}(t)$ coincides with the classical expression in the limit $k_BT\gg\hbar\omega$.
We point out that the difference between $\nu_q^{cor}$ (which arises from the mirror-field correlation) and the classical visibility is negligible even at very low temperatures (at $T=10^{-6}$K and $\omega=2\pi\times 10^5$Hz we have $|\nu_q^{cor}-\nu_c|\le |e^{-2k^2}-1|\sim 0.01$ even when pushing the coupling to $k=0.1$). The parameter $k$ is thus crucial for quantum behavior in such setups, as also discussed in Refs. \cite{penrose2003, kleckner2008}. In the context of a single photon source and within a hybrid framework a similar result was observed in Ref. \cite{lampo2014}.
Moreover, our analysis identifies additional quantum  behavior in $\nu_q^{Kerr}$ due to the quantum-mechanical Kerr non-linear interaction. As Fig.\ref{visibility_qc} shows, while the classical result displays a complete revival after every mechanical period $\tau$, the Kerr nonlinearity lowers the visibility giving rise to a partial revival.

Although the noise in the coherent state has an intrinsic quantum origin, we can bring our classical model closer to the quantum picture. Let us assume our classical coherent field is affected by a gaussian noise \cite{milburn1992}: the field energy in the classical Hamiltonian $H_c$ could be written as $E(\epsilon)=E_0(1-\epsilon)$ where the dimensionless parameter $\epsilon$ is described by the distribution $\mathcal{P}(\epsilon)=1/(\sqrt{2\pi}\Delta)e^{-\frac{\epsilon^2}{2\Delta^2}}$, $\Delta^2$ being the variance. The classical phase in Eq. \eqref{classphase-t} and the intensities in Eq. \eqref{class-int-ab} will now depend on the noise $\epsilon$. By averaging Eq.\eqref{class-int-ab} over the gaussian distribution, the classical visibility (in terms of the optomechanical parameters) is calculated as
\begin{equation}\label{class-visibility-noise}
\tilde{\nu}_c(t)=\nu_c(t)e^{-2k^4N_p(\omega t-\sin\omega t)^2},
\end{equation}
where we used $E_0=\hbar\omega_fN_p$ and $\Delta^2=1/N_p$ to closely compare our gaussian noise with the (poissonian) quantum noise (see Appendix \ref{appendix5} for more details on the derivation of Eq.\eqref{class-visibility-noise}). Therefore, noise in the classical field allows us to exploit the classical kerr-nonlinearity of the phase (see Eq. \eqref{classical-harmonic-phase-t}) to recover a further loss in the classical visibility, which coincides in the limit $k^2\omega t \ll 1$ and large intensities with the quantum result $\nu_q^{Kerr}$ in Eq. \eqref{visibility}. However, while $\nu_q^{Kerr}$ is periodic so to cause revivals, the classical kerr-nonlinearity only lowers the visibility. This further highlights the importance of the parameter $k$ for quantum behavior in optomechanical systems.

We conclude that quantum and classical visibilities display qualitatively the same trend in the current experimental conditions and in order to observe significant deviations ($|\nu_q-\nu_c|\gtrsim 10^{-4}$ within a mechanical period) we need to improve the coupling or the number of photons to $k\gtrsim 10^{-3}$ and $N_p\gtrsim 10^6$ (with all the other parameters as in Fig. \ref{visibility_qc}), independently of the temperature. Indeed, while $\nu_q$ tends to $\nu_c$ in the limit $k^2N_p\ll\bar{n}$ and $k_BT\gg \hbar\omega$, in the same limit, the classical visibility and its quantum counterpart (for a coherent state) coincide with the quantum visibility for a single photon, found in Refs. \cite{penrose2003, kleckner2008}. This entails that the visibility pattern alone is not sufficient to infer non-classicality of the system dynamics. For a quantum interpretation of the results it is essential to have additional assumptions: for instance, in Refs. \cite{penrose2003, kleckner2008} (where the entanglement between the oscillator and the field causes the collapse of visibility) one has to rely on single photon and a ground state mechanical oscillator. 
Classically, on the other hand, the certainty of the mechanical position for zero effective temperature keeps the maximum visibility without a collapse. However, since any small deviation from zero effective temperature does cause the classical visibility to reduce, an unambiguous proof of quantumness requires additional measurements, such as the verification of the entanglement between field and mechanics. 

\section{\label{sec:level4} Conclusions}

Dynamical operations that modify the phase of a system are used in a variety of optomechanical schemes and play a central role in optomechanics to probe the foundations of quantum theory. Here, we studied the classical and quantum nature of such phases, showing that some key features in recent proposals are reproduced classically. In particular, we have seen that the two main peculiarities of the quantum phase are reproduced classically: the nonlinear interaction induced by the mechanical oscillator and its decoupling at certain interaction times. These findings have further allowed us to challenge the quantumness of the interferometric visibility, which has been considered a quantum signature of the system dynamics. While in the common experimental regimes of large photon numbers and small couplings the classical and quantum descriptions mostly coincide, we isolate genuine quantum signatures of the interaction that appear on the phase and the visibility. These signatures might be probed in future optomechanical experiments, even in the weak coupling limit. We finally remark that the classical results found here derive from a fully classical theory in contrast to other approaches using both quantum operators and thermal fields.


\section{\label{sec:level4} Acknowledgements}

The authors wish to thank Carlo Di Franco and Doug Plato for useful discussions. MSK acknowledges the Leverhulme Trust [Project RPG-2014-055], the UK EPRSC (EP/034480/1) and the Royal Society. FA and MSK acknowledge the Marie Curie Actions of the EU's 7$^{\mbox{th}}$ Framework Programme under REA [grant number 317232] for their financial support. IP acknowledges support by the NSF through a grant to ITAMP. CB acknowledges support from the European Commission project RAQUEL (No. 323970); the Austrian Science Fund (FWF) through the Special Research Programme FoQuS, the Doctoral Programme CoQuS and Individual Project (No. 2462).

\appendix

\section{\label{appendix1}From the pulsed to the continuous interaction: phases and Suzuki-Trotter expansion}
In this appendix we find the phase acquired by the optical field in a general pulsed scheme with $\mathcal{N}$ filed-mirror consecutive interactions.

\textit{Quantum scheme.} We define the general displacement operator $\hat{\xi}_{\mathcal{N}}$ corresponding to a loop (in the quantum phase space of the oscillator) shaping a regular polygon of $\mathcal{N}$ sides
\begin{equation}\label{Loop-Polygon}\begin{split}
\hat{\xi}_{\mathcal{N}}=\prod_{j=0}^{\mathcal{N}-1}e^{i\hat{\eta}\left\lbrace\cos(\theta\cdot j)\hat{x}+\sin(\theta\cdot j)\hat{p}\right\rbrace },
\end{split}\end{equation}
where $\theta=2\pi/\mathcal{N}$. Eq. \eqref{Loop-Polygon} can be calculated by using Baker-Campbell-Hausdorff formula \cite{barnett1997} as $\hat{\xi}_\mathcal{N}=e^{i\hat{\Phi}(\hat{\eta},\mathcal{N})}$, where $\Phi(\eta,\mathcal{N})=\frac{1}{4}\eta^2\mathcal{N}\cot(\pi/\mathcal{N})$ without the hat is the area mapped out by the sequence of displacement operations of amplitude $\eta=\langle \hat{\eta}\rangle$ in phase space.
Taking the limit $\mathcal{N}\rightarrow\infty$ in Eq. \eqref{Loop-Polygon} and rescaling $\eta\rightarrow\eta/\mathcal{N}$ we define a \emph{continuous displacement}
\begin{equation}\label{Continuous-Displacement}\begin{split}
\hat{\xi}_{cont}=\lim_{\mathcal{N}\rightarrow\infty}\hat{\xi}_\mathcal{N}=e^{i\frac{\hat{\eta}^2}{4\pi}}\;,
\end{split}\end{equation}
which corresponds to a circle in the phase space with radius $\eta/2\pi$. In the case of the optomechanical interaction we have $\hat{\eta}=\lambda \hat{n}$. Applying the displacement $\hat{\xi}_{\mathcal{N}}$ to the state $\ket{\psi_0}=\ket{\alpha}_f\otimes\ket{\phi(0)}_m$, with $|\phi(0)\rangle_m$ a generic mirror initial state, we measure the mean value of the optical field
\begin{equation}\label{Opticalphase}
\langle \hat{a}\rangle=\langle\psi_0|\hat{\xi}_\mathcal{N}^\dag \hat{a}\;\hat{\xi}_\mathcal{N}|\psi_0\rangle=\alpha\;e^{-N_p(1-\cos 2c)}e^{i(c+N_p\sin 2c)},
\end{equation}
where $c=(\lambda^2/4)\mathcal{N}\cot(\pi/\mathcal{N})$. The first exponential factor of the right hand side represents the change of the size of the amplitude, while the second one gives the change of phase
\begin{equation}\label{Opticalphase-2}
\varphi_q=\frac{\lambda^2}{4}\mathcal{N}\cot\left(\frac{\pi}{\mathcal{N}}\right)+N_p\sin\left[\frac{\lambda^2}{2}\mathcal{N}\cot\left(\frac{\pi}{\mathcal{N}}\right)\right].
\end{equation}
which in the small coupling gives
\begin{equation}\label{Opticalphase-limit}\begin{split}
\varphi_q\simeq &\frac{\lambda^2}{4}\mathcal{N}\cot\left(\frac{\pi}{\mathcal{N}}\right)(1+2N_p)\\
&=\frac{\hbar N_{rt}^2\mathrm{k}_f^2}{m\omega} \mathcal{N}\cot\left(\frac{\pi}{\mathcal{N}}\right)(1+2N_p),
\end{split}\end{equation}
where we used $\kappa=c/(2LN_{rt})$ and $\omega_f=c\mathrm{k}_f$. Having closed polygons in phase space (lasting for an entire period $\tau$) ensures that the phase does not depend on the initial mirror state \cite{bose1997, mancini1997}.

\begin{figure}[h!]
\centering
\includegraphics[scale=0.25]{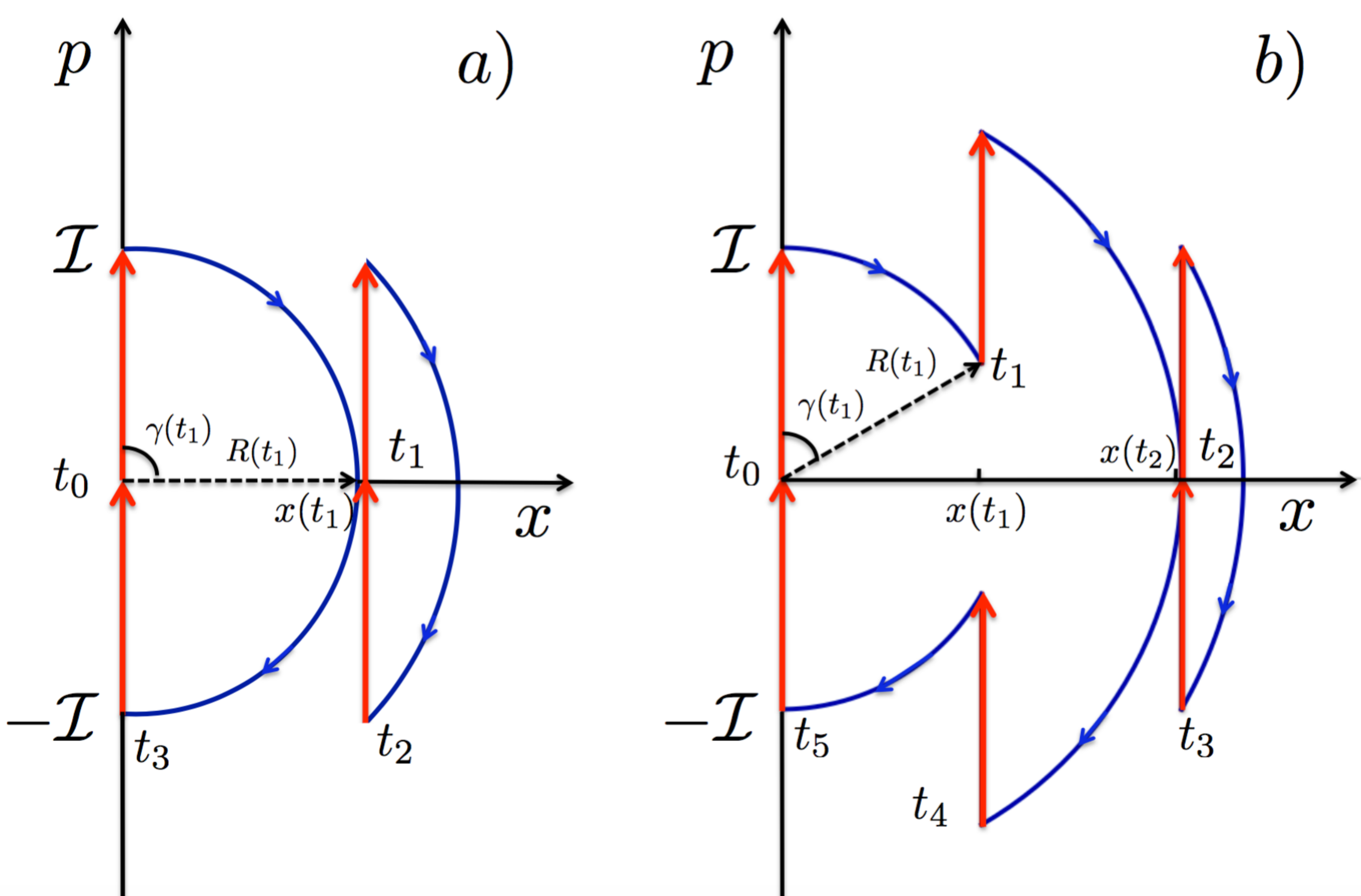}
\caption{Phase space description of the dynamics of the light pulse-mechanical oscillator interaction in the classical picture. a) Four pulse interaction model: the oscillator is assumed at rest at the origin of the phase space. The oscillator gains a momentum $\mathcal{I}$ due to the interaction at $t=t_0$. Then it freely evolves to the maximum amplitude $x(t_1)$ when the second pulse happens, this causes another momentum gain of the oscillator at time $t_1$. At this time the oscillator evolves to $x(t_2)=x(t_1)$ where the third pulse interaction brings its momentum to zero. Now, it evolves to $x(t_3)=0$ where its momentum becomes $-\mathcal{I}$. Finally, the oscillator is brought back to the origin of the phase space by the last pulse-oscillator interaction. b) A similar dynamics is plotted for the six pulse interaction.
}\label{classicalspace}
\end{figure}

\textit{Classical scheme.} From a classical point of view we consider a Fabry-Perot cavity with one massive rigid mirror and one small end mirror that can vibrate in a harmonic potential. The larger rigid cavity mirror has a lower reflectivity than the mechanical mirror that allows the light to enter and exit through this mirror with minimal transmission through the movable mechanical mirror. As a result, the cavity has a finesse $\mathcal{F}$ and when the field enters into the cavity, it is reflected by the movable boundary a number of times equal to the number of round trips inside the cavity, that is $N_{rt}=\mathcal{F}/\pi$. After all these reflections, during which the position of the movable mirror is essentially fixed, the field transfers a momentum $\mathcal{I}$ to the movable mirror (a radiation-pressure kick). The optical field can thus escape the cavity after a time equal to $1/\kappa=(2L/c)N_{rt}$ and then waits in an engineered loop before being initialized again. During consecutive kicks the mirror freely evolves as $x(t)= x(t_0)\cos\omega t+p(t_0)/(m\omega)\sin\omega t$. Following Refs. \cite{rakhmanov2001, rakhmanov2002} for every radiation-pressure kick the field picks up an additional phase due to the movable mirror changing its position. Without loosing generality, we suppose the mirror initially at the origin: at the first kick we have $x(t_0)=0$, $p(t_0)=\mathcal{I}$, and consequently the position evolves as $x(t)= \mathcal{I}/(m\omega)\sin\omega t$ until the second kick. The additional phase shift of the field escaping the cavity after $\mathcal{N}$ light kicks on the mirror, occurring at times $t_j=2j\pi/(\mathcal{N}\omega)$, results $\varphi_{c}=2\mathrm{k}_fN_{rt}\sum_{i=0}^{\mathcal{N}-1}x(t_i)$, where $x(t_i)$ are the classical positions of the mirror at times $t_i$. In Fig.\ref{classicalspace} we show a loop in the classical phase space of the harmonic oscillator in the case of four and six kicks. We remark that the generalization to a generic initial condition is straightforward by simply applying a translation in phase space. The positions that appear $\varphi_{c}$ can be computed through geometric considerations and depicted in the classical phase space of the mirror with polar coordinates [$R(t_i),\vartheta(t_i)$]. At the first kick $R(t_0)=0$ and $\vartheta(t_0)=0$  while, for the consecutive kicks, i.e. $i=1,...,(\mathcal{N}-1)$, we have
\begin{equation}
\begin{split}
R(t_i)&=\sqrt{\zeta^2+2 R(t_{i-1})\zeta\cos{[\vartheta(t_{i-1})]}+R(t_{i-1})^2}\\
\vartheta(t_{i})&=\frac{2\pi}{\mathcal{N}} + \arcsin{\left[\frac{R(t_{i-1})}{R(t_{i})}\sin{[\vartheta(t_{i-1})]}\right]},
\end{split}
\end{equation}
where $\zeta=\mathcal{I}/(m\omega)$ quantifies the classical displacement. Since $x(t_i)=R(t_i)\sin(\vartheta(t_i))$, by numerically solving this recurrence it can be shown that the sum of the oscillator positions corresponds to $(\mathcal{I}/2m\omega)\;\mathcal{N}\cot\left(\pi/\mathcal{N}\right)$. Therefore, we obtain for the classical phase
\begin{equation}\label{classical-harmonic-phase}\begin{split}
\varphi_{c}=\mathrm{k}_fN_{rt}\frac{\mathcal{I}}{m\omega}\mathcal{N}\cot\left(\frac{\pi}{\mathcal{N}}\right).
\end{split}\end{equation}

\textit{Quantum vs Classical phases.}  The momentum transferred at each kick to the movable mirror can be written as $\mathcal{I}=2N_{rt}(E_0/c)$. In order to compare the classical and quantum results, we use $E_0=N_p\hbar\omega_f$. The classical phase shift is thus rephrased as
\begin{equation}\label{classical-harmonic-phase-2}\begin{split}
\varphi_{c}=\frac{2\hbar N_{rt}^2\mathrm{k}_f^2N_{p}}{m\omega} \mathcal{N}\cot\left(\frac{\pi}{\mathcal{N}}\right).
\end{split}\end{equation}
By comparing Eqs. \eqref{Opticalphase-2} and \eqref{classical-harmonic-phase-2} the quantum and classical optical phases generally differ, because Eq.\eqref{Opticalphase-2} holds also for strong coupling regimes: the \emph{$+1$} term in Eq.\eqref{Opticalphase-limit} reveals quantum peculiarities due to the quantization of both field and mechanical oscillator. Nevertheless, for the most common experimental conditions, i.e. small coupling ($\lambda\ll 1$) and strong laser sources ($N_{p}\gg 1$), quantum and classical phases coincide.

\textit{Trotter-Suzuki expansion.} In order to mathematically derive the description of a continuous interaction from the discretized one, we observe that the rescaled limit $\mathcal{N}\rightarrow\infty$ in Eq.\eqref{Loop-Polygon} looks like Trotter's expansion \cite{trotter1959, suzuki1976} for the evolution operator $\hat{U}=e^{-i\hat{H}t/\hbar}$. Indeed, by algebraic manipulations we get
\begin{equation}\label{Ucontinuous}\begin{split}
e^{-\frac{i}{\hbar}\hat{H}t}&=\lim_{\mathcal{N} \to \infty}(e^{-\frac{i}{\hbar}\hat{H}_0t/\mathcal{N}}\; e^{-\frac{i}{\hbar}\hat{H}_It/\mathcal{N}})^\mathcal{N}\\
&=\lim_{\mathcal{N} \to \infty}\prod_{j=0}^{\mathcal{N}-1}(e^{-\frac{i}{\hbar}\hat{H}_0tj/\mathcal{N}}\; e^{-\frac{i}{\hbar}\hat{H}_It/\mathcal{N}}\; e^{-\frac{i}{\hbar}\hat{H}_0tj/\mathcal{N}})\\
&=\lim_{\mathcal{N} \to \infty}\prod_{j=0}^{\mathcal{N}-1}e^{ig_0\hat{n}(\hat{x}\cos\theta_j+\hat{p}\sin\theta_j)\frac{t}{\mathcal{N}}},
\end{split}\end{equation}
where $\theta_j=\omega jt/\mathcal{N}$. Considering an interaction that lasts $\tau$ and bearing in mind that $\kappa=\omega/2\pi$, it then follows that $\hat{n}g_0\tau=\hat{n}\lambda=\hat{\eta}$. We have thus verified that the displacement related to the unitary operator in Eq.\eqref{Ucontinuous} coincides with the circle loop in Eq.\eqref{Continuous-Displacement}. The continuous dynamics can be recovered from the pulsed regime: it is sufficient to keep the light inside the cavity for an interaction time equal to $\tau$ in order to implement a displacement $\hat{\xi}_{cont}$. Also, since we have just established the link between a continuous displacement operation and the unitary operator of the system, the correspondence between classical and quantum phases is expected to hold still in the case of a continuous interaction.

\section{\label{appendix2}Dynamics of the system in case of a continuous interaction}

We will assume field and mirror initially in the state $\ket{\Psi(0)}=\ket{\alpha}_f\otimes\ket{\gamma}_m$ with $\ket{\gamma}_m$  a coherent state of the oscillator.

\textit{Quantum Picture.} The evolution of the state determined by $\hat{U}$ is given by \cite{mancini1997}
\begin{equation}\label{state}\begin{split}
&|\Psi(t)\rangle=e^{-\frac{|\alpha|^2}{2}}\sum_{n=0}^{\infty}\frac{\alpha^n}{\sqrt{n!}}e^{ik^2n^2(\omega t-\sin \omega t)}\\
&\quad\times e^{ikn[\gamma_R\sin \omega t+\gamma_I(1-\cos\omega t)]}|n\rangle_f\otimes \left|\Gamma_n(t)\right\rangle_m
\end{split}\end{equation}
where $\ket{n}_f$ is a Fock state of the cavity field and $\ket{\Gamma_n(t)}_m=\ket{\gamma e^{-i\omega t}+kn(1-e^{-i\omega t})}_m$ the displaced coherent state of the mechanical oscillator. $\gamma_R$ and $\gamma_I$ are respectively the real and imaginary part of $\gamma$.
By tracing out the mechanical degrees of freedom, we obtain the reduced density operator for the field from which we get the mean value of the optical field ($\langle \hat{a}\rangle=\Tr[\hat{a}\hat{\rho}_{f}]$) and the acquired phase at time $t$
\begin{equation}\label{Quantum-Phase}\begin{split}
&\varphi_{q}(\gamma,t)= 2k\left[\gamma_R\sin \omega t+\gamma_I(1-\cos \omega t)\right]\\
&+k^2\left(\omega t-\sin \omega t\right)+N_{p}\sin\left[2k^2(\omega t-\sin \omega t)\right],
\end{split}\end{equation}
For closed loops this result coincides with the one given in the main text.\\
For completeness, we also derive the mean values of the oscillator position and momentum:\begin{equation}\begin{split}\label{quant_motio}
\langle \hat{x}(t)\rangle &= \sqrt{2}\gamma_R\cos\omega t+\sqrt{2}\gamma_I\sin\omega t+\sqrt{2}N_pk(1-\cos\omega t),\\
\langle \hat{p}(t)\rangle &= \sqrt{2}\gamma_I\cos\omega t-\sqrt{2}\gamma_R\sin\omega t+\sqrt{2}N_pk\sin\omega t.
\end{split}
\end{equation}

\textit{Classical Picture.} We verified that in the pulsed regime quantum and classical phases coincide in certain limits. Trotter expansion suggested this to hold also in the continuous case. We show in details that not only this is true for closed loops, but also at every time of the evolution. From a classical perspective by solving the associated Hamilton equations to $H_c$ we obtain the equation of motion
\begin{equation}\label{class_motio}
x(t)= x(0)\cos\omega t+\frac{p(0)}{m\omega}\sin\omega t+ \frac{E_0}{m\omega^2L} (1-\cos\omega t).
\end{equation}
By comparing Eq.\eqref{class_motio} with Eq.\eqref{quant_motio} we see that the dynamics of the mechanical oscillator is harmonic and classical and quantum pictures coincide, even if second order momenta are different. The classical phase shift for a continuous interaction results to be
\begin{equation}\label{Classical-Phase}\begin{split}
&\varphi_{c}(x(0),p(0),t)=2\frac{\omega_f}{c}\frac{1}{d\tilde{\tau}}\int_0^{t}x(\tau)d\tau\\
&=\frac{\omega_f}{L\omega}\left[x(0)\sin\omega t+\frac{p(0)}{m\omega}(1-\cos\omega t)\right]\\
&\quad\quad+\frac{\omega_f}{\omega^3mL^2}E_0(\omega t-\sin\omega t).
\end{split}\end{equation}

\textit{Quantum vs Classical phases.} By using the optomechanical parameters and $E_0=\hbar\omega_f N_p$, we rephrase Eq.\eqref{Classical-Phase} as
\begin{equation}\label{Classical-Phase-2}\begin{split}
&\varphi_c(x(0),p(0),t)=k\sqrt{\frac{2m\omega}{\hbar}}\\
&\quad\times\left[x(0)\sin\omega t+\frac{p(0)}{m\omega}(1-\cos\omega t)\right]\\
&\hspace{4cm}+2N_{p}k^2(\omega t-\sin\omega t).
\end{split}\end{equation}
Classical and quantum phases coincide at every time $t$ for every initial condition in the limit $\lambda\ll 1$ and $N_p\gg 1$. To completely access the comparison, we remark that the initial displaced gaussian quantum state $\ket\gamma_m$ corresponds to the classical boundary conditions $x(0)=\sqrt{2}\gamma_R\sqrt{\hbar/(m\omega)}$ and $p(0)=\sqrt{2}\gamma_I\sqrt{\hbar m\omega}$. This equality between the classical and the quantum result for the phase guarantees the same loss and revival of classical and quantum visibilities due to thermal effect.

\section{\label{appendix3}Semiclassical approach}

We first consider a quantum field and a classical oscillator. The field hamiltonian in a frame rotating at frequency $\omega_f$ can be written as $\hat{H}_f=\epsilon \hat{a}^{\dag}\hat{a}x(t)$ with $x(t)$ the classical equation of motion of the oscillator and $\epsilon=\hbar\omega_f/L$ the resulting coupling constant. If the field is initially in the coherent state $|\alpha\rangle_f$, the field density matrix will read
\begin{equation}\begin{split}\label{rho-semiclass}
\hat{\rho}_f(t)&=e^{-|\alpha|^2}\sum_{n,m}\frac{\alpha^n\alpha^{*m}}{\sqrt{n!m!}}e^{-\frac{i}{\hbar}\epsilon(n-m)\int_0^tx(\tau)d\tau}|n\rangle_f\langle m|
\end{split}\end{equation}
and the mean value of the optical field, which gives us the acquired optical phase, is
\begin{equation}\begin{split}\label{a-semiclass-0}
\langle\hat{a}\rangle&=\alpha e^{-\frac{i}{\hbar}\epsilon\int_0^t x(\tau)d\tau}.
\end{split}\end{equation}
If we model the classical mirror as a harmonic oscillator driven by a constant force $E_0/L$ as in Eq. \eqref{classical-hamiltonian}, we can safely substitute the dynamics in Eq. \eqref{class_motio} into Eq. \eqref{a-semiclass-0} obtaining
\begin{equation}\begin{split}\label{a-semiclass}
\langle\hat{a}\rangle&=\alpha e^{-i\varphi(t)},
\end{split}\end{equation}
where the phase $\varphi(t)$ coincides with the classical phase showed in Eq. \eqref{Classical-Phase}. From Eq.\eqref{a-semiclass} we deduce that when the field is quantized and the oscillator is classical we regain the fully classical result for the phase.
\\
We now show that the same happens for the inverse situation, when the field is described classically and the mirror quantum-mechanically. In this case, the phase acquired by the optical field is given by
\begin{equation}\begin{split}\label{phase-semiclass-2}
\varphi(t)=2\frac{\mathrm{k}_f}{d\tilde{\tau}}\int^{t}_{0}\langle\hat{x}(\tau)\rangle d\tau
\end{split}\end{equation}
i.e. the integral over the interaction time of the mean value of the oscillator position. If we assume the mirror initially in a coherent state $|\tilde{\Psi}(0)\rangle=|\gamma_R+i\gamma_I\rangle$, its evolution under the quantum hamiltonian $\hat{H}_m=\hbar\omega \hat{b}^{\dag}\hat{b}-(E_0/L)\sqrt{\hbar/(2m\omega)}(\hat{b}^{\dag}+\hat{b})$ reads
\begin{equation}\begin{split}\label{psi-mirror-semiclass}
&|\tilde{\Psi} (t)\rangle=e^{ik^2N_p^2(\omega t-\sin\omega t)}\\
&\times e^{i2kN_p[\gamma_I(1-\cos\omega t)+\gamma_R\sin\omega t]}
|\gamma e^{-i\omega t}+kN_p(1-e^{-i\omega t})\rangle,
\end{split}\end{equation}
where we used $kN_p=E_0/(L\omega\sqrt{2\hbar m\omega})$ to express the result in terms of the characteristic optomechanical parameters. It can be verified that the mean value of the position operator given by Eq.\eqref{psi-mirror-semiclass} coincides with the results found in Eqs.\eqref{quant_motio} and \eqref{class_motio} within a fully quantum and/or classical description of the interaction. Hence, the phase acquired by the optical field in Eq. \eqref{phase-semiclass-2} coincides with the classical result reported in Eq. \eqref{Classical-Phase-2}. Again, in terms of optical phase shift a semiclassical description provides the same result of the fully classical one. We can then infer that such a semi-classical description is insufficient to describe all features of the full interaction. Similar considerations can be extended to the visibility.

\section{\label{appendix4}Quantum visibility}

In this section we give all the details on the calculations that lead to the quantum visibility measured in the interferometric scheme depicted in Fig. $2$(a). If the field is initially in a coherent state and the mirror is defined by a thermal state, the density matrix of the system at time $t$ is
\begin{equation}\begin{split}
&\hat{\rho}(t)=e^{-|\alpha|^2}\sum_{m,n}\frac{\alpha^n\alpha^{*m}}{\sqrt{n!m!}}e^{ik^2(\omega t-\sin\omega t)(n^2-m^2)}\\
&\quad\quad\times e^{kn(\gamma \bd-\gamma^*b)}\hat{\rho}_{m}(0)e^{km(\gamma^*b-\gamma\bd)}\ket{n}_f\bra{m}.
\end{split}\end{equation}
By tracing out the mechanical degrees of freedom, we obtain
\begin{equation}\begin{split}
&\hat{\rho}_{f}(t)=e^{-|\alpha|^2}\sum_{m,n}\frac{\alpha^n\alpha^{*m}}{\sqrt{n!m!}}e^{ik^2(n^2-m^2)(\omega t-\sin\omega t)}\\
&\quad\quad\times e^{-k^2(n-m)^2(1-\cos\omega t)(2\bar{n}+1)}
\ket{n}_f\bra{m}.
\end{split}\end{equation}
Michelson interferometry depicted in Fig. $2$(a) corresponds to projecting on quadrature operator eigenstates $\hat{X}_{\phi}=(1/\sqrt{2})[\hat{a}_{out(1)}e^{-i\phi}+\hat{a}_{out(1)}^{\dag}e^{i\phi}]$, where $\hat{a}_{out(1)}$ is the field operator that exits the cavity with the mobile mirror. By computing the mean value $\langle \hat{X}_{\phi}\rangle =\Tr[\hat{X}_{\phi}\hat{\rho}_{f}]$  we find the intensities on the two detectors:
\begin{equation}\begin{split}\label{intensities}
&I^a_b(t)=\frac{I_0}{2}\left( 1 \pm \frac{\langle \hat{X}_{\phi}\rangle}{\sqrt{2}} \right)\\
&=\frac{I_0}{2}\lbrace 1 \mp e^{-\left\lbrace k^2[1-\cos\omega t](2\bar{n}+1)+N_{p}[1-\cos(2k^2(\omega t-\sin\omega t))]\right\rbrace}\\
&\hspace{0.4cm}\times\cos[k^2(\omega t-\sin\omega t)-N_{p}\sin(2k^2(\omega t-\sin\omega t))-\phi]\rbrace.
\end{split}\end{equation}
It is then straightforward to recover the expression in Eq. $(4)$.

\section{\label{appendix5}Classical visibility with noise}
Here, we give more details on the derivation of Eq. \eqref{class-visibility-noise}. Supposing that the energy carried by the field in the classical hamiltonian $H_c$ is subjected to a gaussian noise \cite{milburn1992} which follows the distribution $\mathcal{P}(\epsilon)$, we need to further average the intensity obtaining
\begin{equation}\begin{split}\label{Average-ClassInt-bis}
&\langle I^a_b(t)\rangle=\frac{I_0}{2}\bigg\lbrace1 \pm e^{-\frac{\chi^2}{\beta}(1-\cos\omega t)}e^{-\frac{\omega^2}{\omega_f^2}\chi^4E_0^2\Delta^2(\omega t-\sin\omega t)^2}\\
&\times\bigg[\cos\left(\frac{\omega}{\omega_f}E_0\chi^2(\omega t-\sin\omega t)-\phi\right)-\frac{\omega}{\omega_f}\chi^2E_0\Delta^2\\
&\hspace{0.4cm}\times(\omega t-\sin\omega t)\sin\left(\frac{\omega}{\omega_f}E_0\chi^2(\omega t-\sin\omega t)-\phi\right)\bigg]\bigg\rbrace.\\
\end{split}\end{equation}
By operating through the phase shifter we can make $\phi=\frac{\omega}{\omega_f}E_0\chi^2(\omega t-\sin\omega t)$, and the classical visibility will then read like
\begin{equation}\begin{split}\label{Vis-Full-Class-0-bis}
\tilde{\nu}_c(t)=e^{-\frac{\chi^2}{\beta}(1-\cos\omega t)}e^{-\frac{\omega^2}{\omega_f^2}\chi^4E_0^2\Delta^2(\omega t-\sin\omega t)^2},
\end{split}\end{equation}
which exhibits also a loss due to the kerr-nonlinearity experienced by the classical noisy field.

\textit{Quantum vs Classical visibilities.} Now we rephrase the classical result for the visibility in terms of the characteristic optomechanical parameters
\begin{equation}\begin{split}\label{Vis-Full-Class}
\tilde{\nu}_c(t)=e^{-\frac{2k^2}{\beta\hbar\omega}(1-\cos\omega t)}e^{-2N_p^2k^4\Delta^2(\omega t-\sin\omega t)^2},
\end{split}\end{equation}
where we have used $\chi=\sqrt{2/\hbar\omega}k$ and $E_0=\hbar\omega_fN_p$. We highlight that the field energy distribution $E(\epsilon)=E_0(1-\epsilon)$ is equivalent to the photon distribution $N(\epsilon)=N_p(1-\epsilon)$ which has variance $N_p^2\Delta^2$. Therefore, in order to make the classical noise closer to the poissonian (quantum) noise we set $\Delta^2=1/N_p$, obtaining

\noindent
\begin{equation}\begin{split}\label{Vis-Full-Class-bis}
\tilde{\nu}_c(t)=e^{-\frac{2k^2}{\beta\hbar\omega}(1-\cos\omega t)}e^{-2N_pk^4(\omega t-\sin\omega t)^2},
\end{split}\end{equation}
\vspace{0.2cm}

\noindent
which coincides with the result reported in the main text.\\
We finally underline that having taken the mirror initially at its rest position does not affect the generality of our result in Eq. \eqref{Vis-Full-Class}, indeed we can always reconstruct the interference in Eq. \eqref{Average-ClassInt-bis}  by adapting the phase shifter $\phi$ to cancel out the extra initial contribution coming from Eq.\eqref{Classical-Phase}.

\end{bibunit}

\end{document}